\documentclass[aps,prl,showpacs,floats,10pt,twocolumn]{revtex4}
\usepackage{graphicx}
\usepackage{amsfonts}
\usepackage{times}
\begin{document}
\def\E{{\bf E}}

 \def\Q{{\bf Q}}

 \def\D{{\bf D}}

 \def\r{{\bf r}}

\def\dV{{\; \rm d^3}{\bf r}}

 \def\curl{{{\rm curl}\; }}

\def\grad{{{\rm grad}\; }} 

\def\div{{{\rm div}\; }}

\def\p{{\bf p}}

 \def\rhat{\hat {\bf r}} 

\title{Local Simulation
  Algorithms for Coulomb Interactions} \author{A. C. Maggs and V.
  Rossetto} \affiliation{Laboratoire de Physico-Chime Th\'eorique,
  ESPCI-CNRS, 10 rue Vauquelin, 75231 Paris Cedex 05, France.  }
\date{\today}
\begin{abstract}
  Long ranged electrostatic interactions are time consuming to
  calculate in molecular dynamics and Monte-Carlo simulations. We
  introduce an algorithmic framework for simulating charged particles which
  modifies the dynamics so as to allow equilibration using a {\sl
    local}\/ Hamiltonian. The method introduces an auxiliary field
  with {\sl constrained}\/ dynamics so that the equilibrium
  distribution is determined by the Coulomb interaction. We demonstrate
  the efficiency of the method by simulating a simple, charged lattice gas.
\end{abstract}
\maketitle

The electrostatic interaction between two point charges in a medium
with uniform dielectric constant $\epsilon_0$ varies as
\begin{math}
  e_1 e_2/4\pi \epsilon_0 r
\end{math}.
The large numerical value of this energy together with its long range
are such that it is very often the most costly component in the
simulation of charged condensed matter systems.  Naive evaluation of
the electrostatic energies in molecular dynamics and Monte-Carlo
algorithms leads to inner loops where the summation over all pairs
 takes a time which scales as $O(N^2)$ for a
single step in which all $N$ particles are updated.

Many methods are used to improve this poor scaling: The optimized
Ewald algorithm splits the summation between real and Fourier space
and has a complexity of $O(N^{3/2})$ \cite{ewald,ewald32}. By
interpolation of the charge distribution onto a grid, fast Fourier
transform methods allow a scaling in $O(N \log(N))$ \cite{darden}.
Finally, a popular method in very large simulations is
an expansion of the charge distribution using  hierarchical
multipoles \cite{multipole,greengard}.  The asymptotic
improvement in efficiency comes, however, with great increases in the
complexity of the coding, especially when distributed on
multiprocessor computers. The numerical prefactors in these scaling
laws are uncomfortably high: Despite the great effort put into
optimizing the electrostatic loop, it is found that in the simulation
of a large biomolecule (with $N \sim 10^5$)
 the great majority of the CPU time is still
used in the Coulomb loop \cite{schlick} in even the most sophisticated
numerical codes. 
Most of these ``fast'' methods can only
be used efficiently in molecular dynamics simulations; there are many
occasions where one would like to perform efficient Monte-Carlo
simulations due to the stability and simplicity of the method.

The classical methods for treating charged systems have another
disadvantage, their inability to treat systems with inhomogeneous
dielectric constants. 
Dielectric inhomogeneities have drastic effects on material
properties. For instance the dielectric contrast between water and the
 core of proteins leads to expulsion of counter-ions from a $3A$
thick hydration layer \cite{perutz}.  To treat charging effects in
proteins quite arbitrary, uncontrolled approximations are made
\cite{awful} on effective electrostatic interactions in the vicinity
of a protein in order to reduce interactions to effective pair-wise
additive potentials.  Similarly much work \cite{polyam} has been
performed on the phase structure of charged synthetic polymers while
neglecting the large dielectric contrasts between water based solvents
and oily backbone structures which are surely important in the
discussion of the stability of the necklace structures predicted in these
systems. 
At present the most
promising algorithms are based on the non-local Marcus energy
functional \cite{marcus,marchi}.

This letter introduces a local algorithm with a propagating field ${\bf
  E}$ with purely local dynamics on an interpolating grid; it has a
complexity in $O(N)$ and is elementary to implement.  In contrast to
conventional grid methods we do not solve for all the field variables
at each integration step; we let the field evolve with its own {\sl intrinsic}
 dynamics.  
We were motivated by the observation that
Maxwell's equations, which are local,  produce Coulomb interactions due
to the propagation of a vectorial field.  These dynamic
equations are, as we shall see, not the only dynamic way of generating
the Coulomb interaction. 
  Our method allows a direct, local implementation of dielectric inhomogeneities. 
As a demonstration of the method
we present an explicit implementation of a local Monte-Carlo algorithm
for a charged lattice gas.
We note that  techniques which interpolate  charge  degrees
of freedom onto a lattice are already very well understood;
they form part of standard packages such as Amber \cite{darden}.

We proceed by showing that the Coulomb interaction can be derived
from a {\sl constrained}\/ variational problem. We then show that the
constraint equations are solved locally if we allow electric fields
which have both gradient and rotational degrees of freedom. This
freedom can be used to produced a local Monte-Carlo algorithm.
Finally we present a numerical verification of the method.

The energy of a system of charged particles in a uniform dielectric
background is expressed as a function of the electric field $\E$
\begin{equation}
{\cal U} = \epsilon_0 \int {\E^2 \over 2}  \dV
\label{Energy}
\end{equation}
where the electric field is constrained by Gauss's law
\begin{equation}
\div \E -\rho/\epsilon_0=0
\label{constraint}
\end{equation}

It is known from classical electrostatics that
one solution of  equation (\ref{constraint}) is given by $\E = - \grad \phi$ so that
\begin{math}
\nabla^2 \phi = -\rho/\epsilon_0.
\end{math}  The general solution to the constraint eq.
(\ref{constraint}) is thus
\begin{equation}
\E = -\grad  \phi + \curl \Q
\end{equation}
where $\phi$ is unique to within an additive constant
and  $\Q$ arbitrary.
In Fourier space the electric field can be expressed as
\begin{math}
  \E({\bf k}) = -i {\bf k} \phi + i {\bf k} \wedge \Q
\end{math}.
The second term of this expression is perpendicular to ${\bf k}$ so
that there are two physical degrees of freedom in the $\Q$ field,
corresponding to two independent polarization states. We can consider
that the field is due to a static potential plus transverse photons.

 Let us study
the stationary states to the variational problem posed by eq.
(\ref{Energy}) and eq. (\ref{constraint}) by using a Lagrange
multiplier, with the functional
\begin{equation}
{\cal A} =  \int \left[
 \epsilon_0{\E^2 \over 2} - \lambda({\bf r}) (\epsilon_0 \div  \E - \rho) 
\right]
 \dV \quad,
\end{equation}
implying that
\begin{math}
\E + \grad  \lambda =0 
\end{math}.
The Lagrange multiplier is identical to the static electrostatic
potential, $\phi$ and the minimum energy is
\begin{math}
{\cal U}_{Coulomb} = {\epsilon_0\over 2}\int (\grad \phi)^2   \dV
\label{E0}
\end{math}.
Consider the energy eq.  (\ref{Energy}) for
an arbitrary $\E$ satisfying the constraints then
\begin{eqnarray}
{\cal U} = {\epsilon_0\over 2}\int\left[
 (\grad \phi)^2  + {(\curl \Q)^2}
\right]
 \dV
\label{nocross}
\end{eqnarray}
Cross terms vanish, as is shown by integrating by parts. 

We now turn to the statistical mechanics of a field with the energy of
eq. (\ref{Energy}) constrained by Gauss's law. We do not impose that
the electric field is calculated from a potential.  The partition
function of a fixed set of charges in presence of a fluctuating field
$\E$ is given by
\begin{equation}
{\cal Z}(\{\r\}) = \int {\cal D} \E\, e^{-\int {\beta \epsilon_0 \E^2\over 2 } \dV}\,
 \prod_{\bf r} \delta(\div \E - \rho(\r)/\epsilon_0) \label{Z}
\end{equation}
The argument $\{\r\}$ denotes the fact that the integral over the
particle positions has not yet been performed.  There are two ways of
treating this equation. Either one introduces an integral
representation of the delta function or one notes that the integral
over the field $\E$ decomposes into a (unique) gradient term and a
(non unique) rotation so that
\begin{equation}
  {\cal Z}(\{\r\}) = {e^{-{\beta\epsilon_0\over 2} \int (\grad \phi)^2 \dV}} 
\int {\cal D} \E_t\, e^{{- {\beta\epsilon_0\over2}}\int \E^2 \dV}  \
\label{Z2}
\end{equation}
where ${\cal D} \E_t= \prod_{\bf r}\delta(\div \E ) {\cal D} \E\,$ performs the summation over all the rotational
degrees of freedom of the field described by the potential $\Q$.
  All the dependence on the particle
positions is in the prefactor characterized by the electrostatic
potential $\phi$ found by solving Poisson's equation. This prefactor
gives the Coulomb interaction between the particles.
The remaining integral is independent of the positions of the charges;
rather remarkably integration over the full set of fields allowed by
the constraint multiplies the standard partition function by a simple
constant. This extra factor in the partition function can be ignored.

In the presence of non-uniform dielectric media one proceeds in a
similar manner with the energy
\begin{equation}
{\cal U} = \int {\D^2 \over 2\epsilon(\r)} \dV
\end{equation}
and the constraint
 \begin{math}
   \div \D -\rho =0
\end{math}.
We deduce that the displacement is given by
\begin{math}
  \D = -\epsilon\, \grad \phi + \curl \Q
\end{math}
with 
\begin{math} 
\div (\epsilon\, \grad \phi) = - \rho
\end{math}
and
\begin{equation}
  {\cal Z} (\{\r\}) = {e^{-{\beta\over 2} \int\epsilon(\r) (\grad \phi)^2 \dV} } 
\int {\cal D \E}_t\, e^{{- \int {\beta\over2 \epsilon(\r)}} \E^2 \dV}
\end{equation}
so that
\begin{equation}
{\cal Z}  (\{\r\}) = {\cal Z}_{Coulomb}  (\{\r\}) {\cal Z}_{fluct}  (\{\r\})
\end{equation}
This time the normalization is a function of the distribution of
dielectric inhomogeneities.  When implemented in inhomogeneous media
our treatment leads to potentials which are the sum of the Coulomb and
a fluctuation potential \cite{casimir} which varies as $1/r^6$ for two
widely separated particles.  This term comes from thermally driven
dipole-dipole interactions: Fluctuations in the field produce an
inhomogeneous polarization, ${\bf P}$, of the dielectric background.
This produces an equivalent charge density of $-\div {\bf P}$ which
interacts via Coulomb's law. Such fluctuation potentials are to be
expected from the Lifshitz theory of dielectrics.



We now propose a lattice version of the above equations suitable for
numerically studying the thermodynamics of charged systems. The trick
is to use the arbitrary vector potential $\Q$ to simplify the
calculation of the updated fields after the motion of a charged
particle. We need, also, to sum over all rotational degrees of freedom
of the field in order to calculate statistical weights from the
partition function eq. (\ref{Z}).

\begin{figure}[htb]
  \includegraphics[scale=.4,] {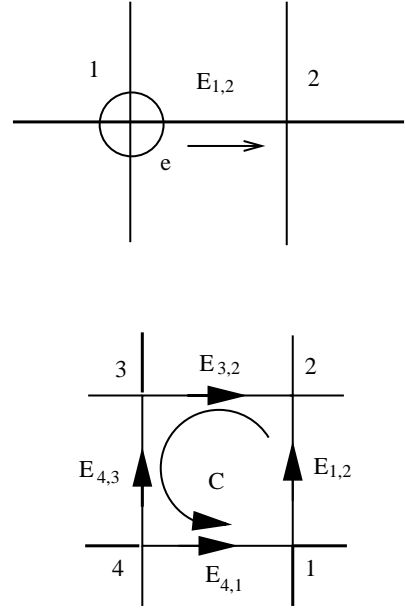}
\caption{Top: A charged particle is present on the left lattice point.
  If the particle is transfered  to the right
  lattice point the constraints  are still satisfied if $E_{1, 2}$
 is modified
  to $E_{1, 2}-e/\epsilon_0$
 on the connecting link where $e$ is the
  charge of the particle.
 Bottom: The four fields associated with a single
  plaquette, $C$, are modified by a rotational motion.
}
\label{conservation}
\end{figure}
The observation that we use in order to implement a local algorithm is
that the constraint equation, eq. (\ref{constraint}) can be updated
{\sl locally}\/ in a system in which charge is conserved. We firstly
reinterpret the constraint in terms of Faraday's concept of
conserved electric flux:
  Consider,
fig.~\ref{conservation} (top),
a  network where the charges are confined to the vertices $\{i\}$ and the field
is associated with links between two sites, $\{i,j\}$.
 Around each lattice point we imagine a cube and
write the constraint in integral form 
\begin{math}
\int \E .\, d {\bf S} =         e_i/\epsilon_0
\label{integconstraint}
\end{math}.
The integral is over the surface of the cube, $e_i$ is the enclosed
charged at the site.
We use the notation $E_{1, 2}$ to denote the total flux leaving $1$
towards $2$; clearly $E_{1, 2}=-E_{2,1}$.

The discretized version of the integral constraint  
is
\begin{math}
 \sum_{j} E_{i, j} = e_i/\epsilon_0
\end{math}
The discretized energy
is given by
\begin{equation}
{\cal U}={\epsilon_0\over 2}\sum_{links} E_{i, j}^2
\end{equation} Here and in what follows we assume that the lattice spacing
is unity.

Start with a system where the constraint is satisfied, fig.~\ref{conservation} (top),
 and displace a charge, $e$, situated on the
leftmost lattice site, $1$, to the rightmost site, $2$. The constraint is
again satisfied at both   sites if the field associated
with the connecting link is updated according to the rule 
$E_{1,2}
\rightarrow E_{1,2} - e/\epsilon_0$.
  This is our Monte-Carlo move
for the particles, involving a correlated update of a single charge
and the field on the link connecting two sites. To update the field
configurations fig.~\ref{conservation} (bottom) we update all the
field values of a single plaquette while conserving the constraint at
each vertex.  In fig.~\ref{conservation} (bottom) $E_{1,2}$ and
$E_{4,1}$ increase by an increment $\Delta$ whereas $E_{4,3}$
and $E_{3,2}$ decrease by $\Delta$ so that at each vertex the sum
of the entering and leaving fields is constant.  It is this last
update that performs the integration over all the rotational degrees
of freedom in the $\E$ field.

The two moves are not quite sufficient to equilibrate a system with
periodic boundary conditions {\sl in all situations}.
 This problem is linked with the solution
\begin{math}
\phi = -{\bar \E . {\bf r}}
\end{math}
or $\E= \bar \E$
of the Laplace equation on a torus where ${\bar \E}$ is an arbitrary constant vector.
Motion of the charges generates fluctuations in ${\bar \E}$, while
 updates such as those in 
fig.~{\ref{conservation}} (bottom) preserve $\bar \E$;
similar phenomena also occur with Maxwell's equations.
In order to be absolutely sure that the algorithm is ergodic
  we introduced a third possible Monte-Carlo step which
consists of a shift in  $\bar \E$.
 By keeping track of the evolution of
$\bar \E$ as the particles move this last update can be efficiently
implemented without destroying the $O(N)$ scaling of the algorithm.
In the largest systems fluctuations in this single mode should
give a small contribution to the thermodynamics if the initial condition
is typical; in such cases this update can be eliminated.

We have performed two initial verifications of the algorithm. Firstly we
randomly placed four  positive and four negative charged particles 
  on a $4 \times 4 \times 4$ 
lattice with periodic boundary conditions.
 We performed field updates using the Metropolis
algorithm at zero temperature in order to anneal the field $\E$. We
then solved for the electric fields using a standard linear algebra
package. The results were identical to within numerical errors.
Annealing  $\bar \E$ was crucial in order to get agreement
between the two methods with frozen charges.
 A second check was then performed with two different implementations of
the Metropolis algorithm now run at several finite temperatures:
36
mutually avoiding, charged particles were distributed in a $6 \times 6
\times 6$ cube with periodic boundary conditions. The dielectric
constant was uniform. In the first simulation the linear solver was
used to calculate the exact interaction energy of the charged
particles at each Monte-Carlo step.  In the second simulation we
implemented the above local algorithm. For the two methods we then
compared the static structure factor for charge-charge
correlations, finding good agreement. 

\begin{figure}[htb]
  \includegraphics[scale=.4,angle=270] {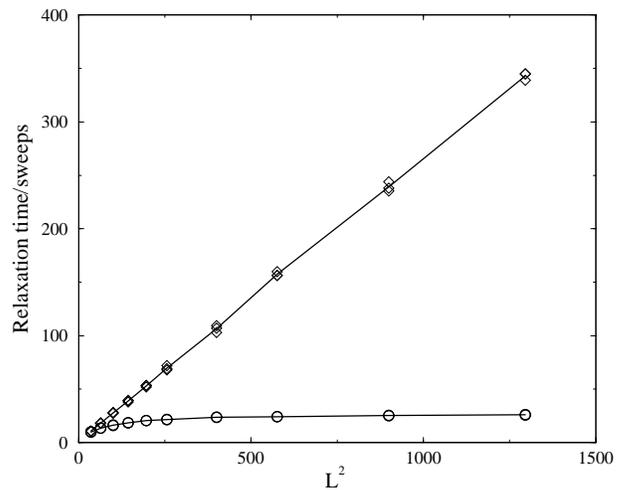}
\caption{
\label{fig2}
Relaxation times in Monte-Carlo sweeps for the 
density-density (top) correlation function
and the charge-charge (bottom) correlation function measured for the mode
${2\pi \over L}(1,0,0)$ plotted as a function of $L^2$.
The top curve shows scaling compatible with simple diffusion.
 The bottom curve saturates for large systems sizes. The multiple points
for each $L$ correspond to  three independent determinations of the
relaxation time. System sizes between $L=6$ and $L=36$ with $L^3/4$ 
charged particles. $6\times 10^5$ sweeps per simulation for a total
simulation time of three days on a AMD Athlon computer.
}
\end{figure}

We  verified the efficiency of the algorithm by determining
the autocorrelation time of the slowest density and charge modes
in a cube of dimension $L$ using the method described in \cite{acm2},
 fig. (\ref{fig2}).
We plot the relaxation time in
Monte-Carlo sweeps; during each sweep
an average of one Monte-Carlo trial is performed for each
degree of freedom in the simulation.
As expected in a Monte-Carlo algorithm dominated by diffusive motion
the slowest mode relaxes in a time which varies as $L^2$.
There is no anomalous or critical slowing down due to the 
coupling between the particles and the electric field.
The saturation of the relaxation time 
for charge fluctuations
at large $L$
is a consequence of screening;
 charge fluctuations relax by diffusion over the Debye length.

We have shown that the thermodynamics of charged systems can be
simulated locally by introducing a propagating vector field so that
particles interact via retarded, {\sl diffusing}\/ fields. The dynamic
properties of the system are strongly modified but by construction
thermodynamics is an invariant of the propagation dynamics.
Our treatment of the dynamics of the field $\E$ is similar to the
Coulomb or radiation gauge in classical electrodynamics: Normally one writes that
\begin{equation}
\E = -\grad \phi - {\partial {\bf A} \over \partial t}
\end{equation}
where in the Coulomb gauge $\div {\bf A} =0$. 
There is a rather close analogy between our dynamic scheme of solving
for the constraint equations with certain methods of quantization in
the Coulomb gauge: In Dirac's quantization of the electrodynamic field
Gauss's law is a {\sl weak}\/ identity 
\cite{dirac} dependent on the choice of the initial wave function.
 In our simulation
Gauss's law is the result of a restricted choice of possible moves in
the Monte-Carlo algorithm together with a special initial condition.


What are the advantages of the present method over direct integration
of Maxwell's equations which are also an example of an $O(N)$ algorithm 
\footnote{We thank the reviewer for this remark}?
  Monte-Carlo algorithms  are particularly easy 
to implement and have good stability with large step sizes.
In addition, we
have checked that the fast equilibration of the electric degrees of freedom
in fig. (2) allows one to perform field updates far more rarely than  particle
updates leading to additional acceleration of the algorithm. Such multiple time step
ideas have been applied to conventional electrostatic solvers but
in molecular dynamics are sometimes prone to numerical instabilities \cite{schlick}.


In the implementation of the algorithm we were inspired by recent work
on hydrodynamic interactions (varying as $1/r$) via a
Lattice-Boltzmann algorithm in simulations of polymer solutions
\cite{ralf}.  Another analogous problem to ours has been treated by
Car and Parrinello who have shown \cite{car} that introduction of a
fictitious dynamics leads to much improved efficiencies in solving for
constraints; our work differs
in that the constraint of Gauss's law is solved {\sl exactly}\/ at
each simulation step whereas the Car-Parrinello algorithm leads to a
simulated annealing solution of the constraint equations (\ref{constraint}).

Finally let us note that other
functionals do exist for the electric potential in the presence of
sources \cite{jackson}.  In particular the functional 
\begin{equation}
\int \left[
{\epsilon_0 \over 2}
(\nabla \phi)^2 - \rho \phi
\right]
 \dV
\end{equation}
 seems, at first sight particularly 
simple and attractive. 
Unfortunately, the minimum of this functional is
{\sl minus} the correct electrostatic energy. 
It can not be used as
a functional for {\sl both} the field evolution {\sl and} the particle
motion.
Application of the algorithm to large atomistic systems remain to be
tested, but our method provides an  alternative to
existing treatments of Coulomb interactions.

 We would like to thank R. Everaers for many crucial
discussions in the formulation of this work, in particular for his
remarks on the importance of dielectric effects.



\bibliography{mc}
\end{document}